\documentclass[12pt,a4paper]{article}
\usepackage[english]{babel}
\usepackage{graphics}
\def\bk{{\bf k}}
\def\bp{{\bf p}}
\def\br{{\bf r}}

\def\b\sigma{\mbox{\boldmath $\sigma$\unboldmath}}

\def\b\xi{\mbox{\boldmath $\xi$\unboldmath}}

\def\cl{{\cal L}}
\def\bX{{\bf X}}
\def\bR{{\bf R}}
\def\({\left(}
\def\){\right)}
\newcommand{\be}{\begin{equation}}
\newcommand{\ee}{\end{equation}}
\newcommand{\beq}{\begin{eqnarray}}
\newcommand{\eeq}{\end{eqnarray}}

\title{ \bf {The Bose gas beyond mean field}}
\author{Philippe A. Martin\\
Institute of Theoretical Physics\\ Swiss Federal Institute for Technology
Lausanne\\ CH-1015, Lausanne EPFL, Switzerland\\ \vspace{3mm}\\
Jaros\l aw Piasecki\\
Institute of Theoretical Physics, University of Warsaw, Ho\.za 69
\\00 681 Warsaw, Poland}

\date{\today}

\begin{document}

\maketitle

\begin{abstract}
We study a homogeneous Bose gas with purely repulsive forces.
Using the Kac scaling of the binary potential we  derive analytically
the form of the thermodynamic functions of the gas for small but finite
values of the scaling parameter in the low density regime. 
In this way we determine dominant corrections 
to the mean-field theory. It turns out that repulsive forces increase the 
pressure at fixed density and  decrease the density at given 
chemical potential (the temperature is kept constant).
They also flatten the Bose momentum distribution. However,
the present analysis cannot be extended to the region where the mean-field 
theory predicts the appearence of condensate.
\end{abstract}
\vskip 1truecm

\section{Introduction}

Whereas the mean-field theory of an interacting Bose gas is now well
understood, going beyond the mean-field description in a systematic way
presents still an open problem (see \cite{MP} and references given therein).
A natural tool for such a study is provided by the Kac scaling of the
binary potential $V(r)$. The scaled potential
\be
V_{\gamma}(r) = \gamma^{3}V(\gamma r) \label{1}
\ee
becomes weak and long-range when the dimensionless parameter $\gamma$
approaches zero. However, the integrated potential energy
\be
a = \int d{\bf r}V(r) = \int d{\bf r} V_{\gamma}(r) < \infty
\label{2}
\ee
is $\gamma$-independent, and thus remains constant.
The so-called van der Waals limit (i.e. the thermodynamic limit followed
by $\gamma \to 0$) permits then to derive the mean-field effects corresponding
to the potential $V(r)$. They depend on the mean potential energy $a\rho$,
where $\rho$ is the number density of particles.
For example, it has been shown that in the case of repulsive forces the van der
Waals limit of the free energy density $f$ yields the mean-field formula
\be
f_{mf} = f_{0} + \frac{a}{2}\rho^{2}  \label{3}
\ee
where $f_{0}$ refers to the reference system without interaction $V(r)$.
The formula (\ref{3}) holds both for classical and for quantum gases
\cite{LP},\cite{L}.
In the theory of classical fluids one could go even further: the
corrections to the zero order mean-field results have been derived,
exhibiting the role of fluctuations for small but finite values of $\gamma$
\cite{H}.

In the present paper we perform an analogous analysis for
a homogeneous Bose gas, continuing our previous study \cite{MP}. 
Our object is thus to determine and to investigate
the leading  corrections to the mean-field theory predictions
for equilibrium properties of a gas composed of identical bosons interacting
via purely repulsive forces.
The rigorous part of our analysis will be restricted the region of thermodynamic parameters in which
quantum Mayer series (virial expansions) converge uniformly with respect
to the small parameter $\gamma$.  In fact, under the assumption that
both the potential and its Fourier transform are
non-negative
\be
V(r) \ge 0, \;\;\;\;\; \hat{V}(k) =
\int d{\bf r}\;{\rm exp}(-i{\bf k}\cdot{\bf r})V(r) \ge 0, \;\;\;   \label{4}
\ee
we can directly apply here the main results of \cite{MP}.
In \cite{MP} the Bose gas is represented as a classical-like system of random
polymers by using the Feynman-Kac path integral formulation of the Gibbs
weight together with a cycle decomposition of the permutation group.
Applying standard Mayer graph summation technique
it has been shown therein that  the density $\rho$ and
the chemical potential $\mu$ of the Bose gas were related by a self-consistent
relation
\be
\rho (\mu) = F(\mu - a\rho (\mu ))      \label{5}
\ee
The function $F$ is defined by a convergent series of multiconnected
graphs whenever the chemical potential is sufficiently negative, and the
convergence is uniform with respect to the scaling parameter $\gamma$.
In this regime, and for $\gamma \ll 1$ the function $F$ for a scaled
potential (\ref{1}) takes the asymptotic form
\be
F^{\gamma}(\nu) = F^{(0)}_{tree}(\nu) + \gamma^{3} F^{(1)}(\nu) +
o(\gamma^{3})   \label{6}
\ee
Here $F^{(0)}_{tree}(\nu)$ represents the sum of zero order contributions from
the tree diagrams. It turns out that
\be
F^{(0)}_{tree}(\nu) = \rho_{0}(\nu)   \label{7}
\ee
where $\rho_{0}$ denotes the perfect gas density.
The limit $\gamma \to 0$ yields thus the self-consistent mean field equation
\be
\rho_{mf}(\mu) = \rho_{0}(\mu - a\rho_{mf}(\mu))
\label{8}
\ee
Equation (\ref{6}) implies that the dominant corrections to equation
(\ref{8}) are of the order $\gamma^{3}$. The correcting term
\be
F^{(1)}(\nu)= F_{tree}^{(1)}(\nu) +F_{ring}^{(1)}(\nu)
\label{8a}
\ee
results from the
summation of the tree diagrams and  from the summation of the ring diagrams.
In Section 2 the form of the functions $F^{(1)}_{tree}(\nu)$ and
$F^{(1)}_{ring}(\nu)$ is derived. The corresponding
equation of state of the Bose gas is presented and analyzed in Section 3.
It is remarkable that the dominant correction to the pressure has the same
structure as that found by Hemmer for a classical fluid \cite{H}. 
Under the additional assumption that the formula remains valid up to a neighborhood
of the critical free density, we observe that the mean-field theory always underestimates
the pressure in this region.
We study then the one-particle reduced density operator
displaying the nature of the momentum distribution (Section 4).
Concluding remarks are presented in Section 5.

\section{Self-consistent equation beyond mean-field}

The function
\be
F^{\gamma}_{tree}(\nu) = \frac{1}{(2\pi\lambda^{2})^{3/2}}\sum_{q=1}^{\infty}
\frac{{\rm exp}(\beta\nu q)}{q^{3/2}}\kappa^{\gamma} (q)        \label{9}
\ee
when used in the self-consistent equation (\ref{5}) yields the term
corresponding to the summation of the tree diagrams (section V.A in
\cite{MP}) . Here
\be
\kappa^{\gamma}(q) = < {\rm exp}(-\beta U_{\gamma}) >_{q} \label{10}
\ee
is the partition function of a single closed $q-$polymer, $\beta = 1/k_{B}T$,
$T$ is the temperature, $k_{B}$ the Boltzmann constant,
$q$ the number of particles in the polymer and
\be
\lambda=\hbar\sqrt{\frac{\beta}{m}}
\label{10a}
\ee
is the de Broglie thermal wave length.
The asymptotic $\gamma$-expansion of the scaled self-energy $U_{\gamma}$ of
the polymer
follows from the definition given in \cite{MP}
\be
U_{\gamma} = \gamma^{3}\frac{q(q-1)}{2}V(0) +  o(\gamma^{3}),\;\;\;
\gamma\to 0   \label{11}
\ee
so that
\be
\kappa^{\gamma}(q) = 1 -
 \gamma^{3}\frac{q(q-1)}{2}\beta V(0) +  o(\gamma^{3}) \label{12}
\ee
Inserting (\ref{12}) into (\ref{9}) we get
\be
F^{\gamma}_{tree}(\nu ) = \rho_{0}(\nu ) +
 \gamma^{3}F^{(1)}_{tree}(\nu) +  o(\gamma^{3})  \label{13}
\ee
where
\be
F^{(1)}_{tree}(\nu) = \frac{\beta V(0)}{2}\left( f^{(1)}(\nu)-f^{(2)}
(\nu)\right)     \label{14}
\ee
We have adopted the notation
\footnote{In \cite{MP} $f_{\gamma}^{(k)}(\nu)$
designates the $\gamma $-dependent function, including the vertex
contribution $\kappa^{\gamma}(q)$. Here we have set $\gamma = 0$,
and $\kappa^{\gamma}(q)=1$, which yields the definition (\ref{15}).}
\be
f^{(k)}(\nu) = \frac{1}{\beta^{k}}\frac{\partial^{k}}{\partial \nu^{k}}
\rho_{0}(\nu) = \frac{1}{(2\pi\lambda^{2})^{3/2}}
\sum_{q=1}^{\infty}\frac{q^{k}}{q^{3/2}}{\rm exp}(\beta q \nu ) \label{15}
\ee

The dominant contribution from resummation of the ring diagrams
$\gamma^{3}F^{(1)}_{ring}(\nu)$
has been also calculated in \cite{MP} (see eq. (70); it is derived here in
the appendix for the sake of convenience, see eq.(\ref{A.18})). It reads
\be
F^{(1)}_{ring}(\nu) = \frac{1}{2}f^{(2)}(\nu )
\int \frac{d{\bf k}}{(2\pi )^{3}}\;
\frac{[\beta \hat{V}(k)]^{2}f^{(1)}(\nu)}{1+\beta \hat{V}(k)f^{(1)}(\nu)}
\label{16}
\ee
Using the definition (\ref{15}) and denoting respectively by $\rho_{0}'$ and by
$\rho_{0}''$ the first- and the second-order derivative of the perfect gas
density with respect to the chemical potential,   we rewrite
equations (\ref{14}),
(\ref{16}) in the form
\be
F^{(1)}_{tree}(\nu) = \frac{ V(0)}{2}\rho'_{0}(\nu)-\frac{ V(0)}{2\beta }
\rho''_{0}(\nu)
\label{17}
\ee
\be
F^{(1)}_{ring}(\nu) = \frac{1}{2}f^{(2)}(\nu )
\int \frac{d{\bf k}}{(2\pi )^{3}}
\beta \hat{V}(k) \left[1- \frac{1}{1+\beta \hat{V}(k)f^{(1)}(\nu)}\right]
\label{18}
\ee
\[ = \frac{ V(0)}{2\beta }\rho''_{0}(\nu) -
\frac{1}{2\beta}\frac{\partial}{\partial \nu}\int \frac{d{\bf k}}{(2\pi )^{3}}
\log [1 + \hat{V}(k)\rho_{0}'(\nu)] \]
The dominant correction to the mean field
form $\rho_{0}$ of the function $F$ is represented in equation (\ref{8a}) by
a term of the order $\gamma^{3}$ involving the sum of the tree and of the ring
contributions. Adding up (\ref{17}) and (\ref{18})
we eventually find that the function $F^{(1)}(\nu)$ takes the form of the
derivative
\be
F^{(1)}(\nu) = F^{(1)}_{tree}(\nu) + F^{(1)}_{ring}(\nu) =
\frac{\partial g(\nu)}{\partial \nu}
  \label{19}
\ee
where
\be
g(\nu) = \frac{ V(0)}{2}\rho_{0}(\nu)
-  \frac{1}{2\beta}\int \frac{d{\bf k}}{(2\pi )^{3}}
\log [1 + \hat{V}(k)\rho_{0}'(\nu)]   \label{199}
\ee
Having derived the form of $F^{(1)}(\nu)$ we can analyze now the
thermodynamic properties of the Bose gas including the corrections
of the order $\gamma^3$.

\section{Equation of state beyond mean-field}

\subsection{Pressure at order $\gamma^{3}$}

The grand canonical pressure $P(\mu)$ satisfies the thermodynamic relation
\be
\frac{\partial P(\mu)}{\partial \mu} = \rho (\mu) \label{27}
\ee
Considering the chemical potential $\mu$ as a function of
density $\mu = \mu (\rho)$ we get from (\ref{27}) the formula
\be
P(\rho ) = \int^{\rho}_{0} d\sigma \; \sigma
\frac{\partial \mu (\sigma)}{\partial \sigma}
\label{28}
\ee
The $\gamma$-expansion of the chemical potential reads
\be
\mu(\rho) = \mu_{mf}(\rho) + \gamma^{3}\mu^{(1)}(\rho )+ o(\gamma^{3})
\label{29}
\ee
where
\[  \mu_{mf}(\rho) = \mu_{0}(\rho) + a\rho \]
and $\mu_{0}(\rho)=\partial f_{0}(\rho)/\partial\rho$ is the chemical potential
of the perfect gas (compare with equation (\ref{3})).
Up to the terms of order $\gamma^{3}$ the pressure is thus given by
\be
P(\rho ) = P_{0}(\rho) + \frac{a}{2}\rho^{2} + \gamma^{3}P^{(1)}(\rho)+
o(\gamma^{3})
\label{30}
\ee
where
\be
P^{(1)}(\rho) = \int^{\rho}_{0} d\sigma\;
\sigma \frac{\partial \mu^{(1)}(\sigma)}{\partial \sigma }   \label{31}
\ee
The correction $\mu^{(1)}(\rho)$ can be readily determined from the
self-consistent equation (\ref{5}). Indeed, up to terms of order $\gamma^{3}$
\be
\rho = \rho_{0}[\mu_{mf}(\rho ) - a\rho + \gamma^{3}\mu^{(1)}(\rho)]
       + \gamma^{3}F^{(1)}(\mu_{mf}(\rho)-a\rho)  \label{32}
\ee
As $\mu_{mf}(\rho ) - a\rho = \mu_{0}(\rho)$, upon further expanding of the
first term on the right-hand side of (\ref{32}) one finds
\be
\rho =   \rho_{0}[\mu_{0}(\rho)] + \gamma^{3}
[ \rho'_{0}(\mu_{0}(\rho ))\; \mu^{(1)}(\rho)+ F^{1}(\mu_{0}(\rho)) ]
\label{33}
\ee
The identity $\rho_{0}(\mu_{0}(\rho)) = \rho $ implies thus the relation
\be
\mu^{(1)}(\rho) =
- \frac{F^{(1)}(\mu_{0}(\rho))}{\rho'_{0}(\mu_{0}(\rho ))} \label{34}
\ee
Moreover, since $F^{(1)}(\mu)$ is the derivative of the function $g(\mu)$
defined in (\ref{199}), we have
\be
\mu^{(1)}(\rho)= -\frac{\partial \mu_{0}(\rho )}{\partial \rho}
\left[ \frac{\partial g(\mu)}{\partial \mu}\right]_{\mu = \mu_{0}(\rho)}
= -\frac{\partial g(\mu_{0}(\rho))}{\partial \rho}
\label{35}
\ee
Upon inserting (\ref{35}) into (\ref{31}) we get the formula
\be
P^{(1)}(\rho) = -\int^{\rho}_{0} d\sigma \;
\sigma \frac{\partial^{2} g(\mu_{0}(\sigma))}{\partial \sigma^{2}}
= \left( -\rho\frac{\partial}{\partial \rho} + 1 \right)g(\mu_{0}(\rho))
\label{36}
\ee
where the second equality follows from integration by parts and the fact that
$g(\mu_{0}(\rho))|_{\rho =0} = g(\mu)|_{\mu =-\infty} =0$. Using the explicit
form (\ref{199}) of $g(\mu)$ we arrive at the final formula
\be
P^{(1)}(\rho) = \frac{1}{2\beta}
\left( \rho\frac{\partial}{\partial \rho}-1\right)
\int \frac{d{\bf k}}{(2\pi )^{3}}\log [1 + \hat{V}(k)\chi_{0}(\rho)]
\label{37}
\ee
where
\[ \chi_{0}(\rho) = \rho'_{0}(\mu_{0}(\rho)) =
\rho \left( \frac{\partial P_{0}}{\partial \rho}\right)^{-1} \]
denotes the compressibility of the perfect Bose gas.

  Equation (\ref{37}) involves not only the thermodynamic functions
characterizing the perfect gas but also the shape of the binary potential.
It permits thus to write down the equation of state
(\ref{30}) of the interacting Bose gas beyond the mean-field theory.
Remarkably enough,  the additional pressure term $P^{(1)}(\rho)$
representing the effect of fluctuations around the mean field has exactly
the same structure as the analogous term derived by Hemmer for classical
fluids (see \cite{H} eq. (55)).

 \subsection{Low density behaviour}

It is interesting to analyze in more detail the
formula (\ref{37}) in the low density limit,
where the compressibility $\chi_{0}(\rho)$ approaches zero
\be
\chi_{0}(\rho) = \beta \rho  + \frac{\beta}{2}(\pi \lambda^{2})^{3/2}\rho^{2}
+ ...  , \;\;\;\;\;\;  \rho \to 0 \label{38}
\ee
The expansion of the logarithm in (\ref{37}) yields
\be
P^{(1)}(\rho) = \frac{1}{2\beta}\left( \rho\frac{\partial}
{\partial \rho} - 1 \right)\left[ V(0)\chi_{0}(\rho) -
\frac{1}{2} \int \frac{d{\bf k}}{(2\pi )^{3}}[\hat{V}(k)
\chi_{0}(\rho)]^{2}  + ... \right]
\label{38a}
\ee
Inserting here (\ref{38}) we find
\be
P^{(1)}(\rho) = \frac{\rho^{2}}{4}\left[ V(0)\pi ^{3/2} \lambda^{3}
- \beta \int \frac{d{\bf k}}{(2\pi )^{3}}[\hat{V}(k)]^{2} \right]
\label{39}
\ee

The term proportional to $\lambda^{3} \sim T^{-3/2}$ reflects the effect
of quantum statistics whereas the term proportional to $\beta = 1/k_{B}T$
is of classical type, not involvoing the Planck constant. The
repulsive potential $V(0)>0$ tends to increase the
pressure, but the negative classical term acts in the opposite direction.
Clearly, the lower the temperature the more important is the Bose statistics.

Notice that the low density equation of state maintains the mean-field
form
\be
P(\rho) = P_{0}(\rho) + \frac{1}{2}a_{\gamma}\rho^{2}  \label{40}
\ee
but now with a $\gamma$-dependent constant
\[ a_{\gamma} = a + \gamma^{3}\frac{1}{2}
\left[ V(0)\pi ^{3/2}\lambda^{3}
- \beta \int \frac{d{\bf k}}{(2\pi )^{3}}[\hat{V}(k)]^{2} \right] \]
Finally, equation (\ref{40}) should be supplemented with the low density
expansion of the perfect gas pressure
\be
P_{0}(\rho) = \frac{1}{\beta}\left[ \rho - \frac{\rho^{2}}{2}
\pi^{3/2}\lambda^{3} + ...  \right]  \label{41}
\ee
Then equation (\ref{40}) yields the second virial coefficient
at the order $\gamma^{3}$.

\subsection{Critical region}

At this point we make the working hypothesis that the validity of the
formula (\ref{37}) for $P^{(1)}(\rho)$
extends from low density up to the critical density $\rho_{0,c}$ of the
free gas provided that $\gamma$ is small enough
when $\rho$ is close to $\rho_{0,c}$. This is a plausible assumption if in
the range $0\leq \rho<\rho_{0,c}$, at fixed temperature $T$,
the system does not undergo other phase transitions (solidification,
liquefaction).
At $\rho_{0,c}$, the compressiblilty of the free gas is known to diverge as
\be
\chi_{0}(\rho)\sim \frac{c}{\rho_{0,c}-\rho}
\label{40a}
\ee
where $c=1.086 \;\beta \rho_{0,c}^{2}$ \cite{Diu}. For $\rho$ close to
$\rho_{0,c}$ we can write
\beq
P^{(1)}(\rho) &\sim& \frac{1}{2\beta}
\left( \rho\frac{\partial}{\partial \rho}-1\right)
\int \frac{d{\bf k}}{(2\pi )^{3}}\log \left[1 +
\frac{c\hat{V}(k)}{\rho_{0,c}-\rho}\right]\nonumber\\
&\sim&\frac{\rho_{0,c}}{2\beta (\rho_{0,c}-\rho)}\int \frac{d{\bf k}}{(2\pi
)^{3}}\frac{c\hat{V}(k)}{\rho_{0,c}-\rho+c\hat{V}(k)}
\label{41}
\\
&-&\frac{1}{2\beta}\int \frac{d{\bf k}}{(2\pi )^{3}}\log \left[1 +
\frac{c\hat{V}(k)}{\rho_{0,c}-\rho}\right].
\label{42}
\eeq
One sees that $P^{(1)}(\rho)\to \infty$ as $\rho\to\rho_{0,c}$, the
positive term (\ref{41}) being the most divergent one.
Thus, choosing $\rho$ close to
$\rho_{0,c}$ and $\gamma^{3}$ sufficiently small, the pressure correction
$\gamma^{3}P^{(1)}(\rho)$
to the mean field can be made positive. We conclude that fluctuations
beyond mean
field always tend to increase the pressure in the vicinity of $\rho_{0,c}$.

The nature of the divergence depends on the behaviour of $\hat{V}(k)$ as
$k\to \infty$. As an example we consider a power-like decay
$\hat{V}(k)\sim bk^{-\eta},\;k\to\infty,\;\eta>3$.
Then the strongest divergence comes from the large values of $k$ in
the integral (\ref{41}). Choosing $k_{0}$ sufficiently large we find
\beq
&&\frac{\rho_{0,c}}{2\beta (\rho_{0,c}-\rho)}\int_{k_{0}}^{\infty}
\frac{dk}{2\pi^{2}
}k^{2}\frac{bc}{k^{\eta}(\rho_{0,c}-\rho)+bc}\nonumber\\
&\sim& \(\frac{1}{(\rho_{0,c}-\rho)^{1+3/\eta}}\)
\frac{\rho_{0,c}}{8\beta\pi^{2}}\int_{0}^{\infty}duu^{2}\frac{bc}{u^{\eta}+b
c}.
\label{43}
\eeq
Since the second integral (\ref{42}) behaves as
$(\rho_{0,c}-\rho)^{-(3/\eta)}$, we see that the
pressure correction diverges as $(\rho_{0,c}-\rho)^{-(1+3/\eta)},
\rho\to\rho_{0,c}$.
An illustration corresponding to the choice
\beq
V(r) = \frac{a}{8\pi r_{0}^3}\exp(-r/r_{0}), \;\;\;\;  \hat{V}(k) =
\frac{a}{(1 + k^2r_{0}^2)^2}
\label{F1}
\eeq
\[ \gamma^3\pi (\lambda/r_{0})^{3}=1, \quad
a\beta/(2\pi\lambda^{2})^{3/2}=0.1 \]
is presented in Fig.1.
\begin{figure}
\includegraphics{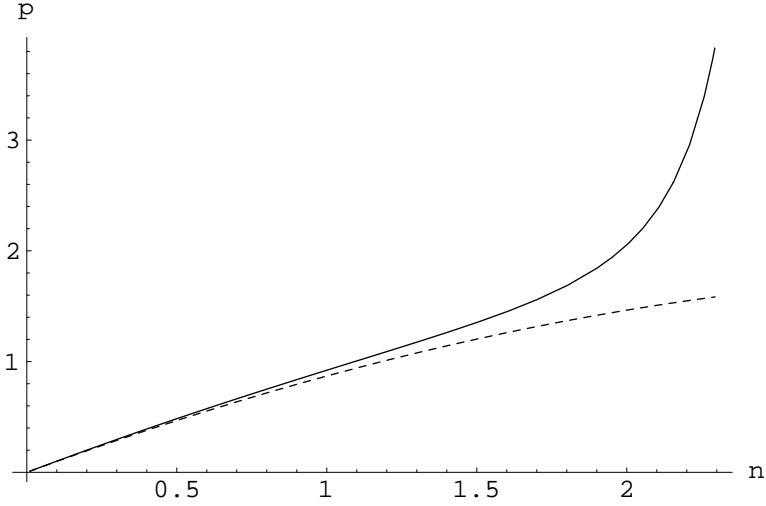}
\caption{Dimensionless pressure (\ref{30}): $ p = P(2\pi
\lambda^2)^{3/2}\beta $ as function of
density $n=\rho (2\pi \lambda^2)^{3/2}$ for the binary interaction
(\ref{F1}). The pressure correction diverges
at $n=n_{0,c}=2.612$. The dashed line
represents the mean-field prediction which underestimates the pressure.}
\end{figure}

\section{One-body density matrix at order $\gamma^{3}$}

The grand canonical one-body density matrix $R$ for a Bose gas in volume
$\Lambda$ is defined by
its configurational matrix elements
\beq
\!\!\!\!&&\langle \br |R(\mu)|\br'\rangle=\nonumber\\
&&\frac{1}{\Xi_{\Lambda}}\sum_{N=1}^{\infty}\frac{e^{\beta\mu N}}{(N-1)\;!}
\int_{\Lambda}d\br_{2}\cdots
\int_{\Lambda}d\br_{N}\langle \br,\br_{2},\ldots,\br_{N}|e^{-\beta
H_{N,\Lambda}}|\br',\br_{2},\ldots,\br_{N}\rangle_{sym}\nonumber\\
\label{4.1}
\eeq
with $H_{N,\Lambda}$ the $N$-particle Hamiltonian and the matrix elements
taken in the space of
symmetrized wave functions. In the low density regime it admits the
classical-like representation (equations (\ref{A.2}) and (\ref{A.5})
of the Appendix) in terms of an open polymer
immersed in a gas of closed polymers. In the thermodynamic limit Mayer
graphs resummations enable to determine $\langle \br
|R(\mu)|\br'\rangle$  from the formula (see Appendix, eq. (\ref{A.7}))
\be
\langle \br |R(\mu)|\br'\rangle = F_{\br,\br'}(\mu-a\rho(\mu))
\label{4.2}
\ee
where, as in (\ref{5}),  $F_{\br,\br'}(\nu)$ is the sum of multiply
connected graphs with a root point labelled by an open
polymer.
It reduces to $F(\nu)$ as $\br=\br'$, so the diagonal part $\langle \br
|R(\mu)|\br\rangle=\rho(\mu)$ satisfies
the self-consistent equation (\ref{5}).
Because of the invariance under translations and under rotations
$F_{\br,\br'}(\nu)=F(|\br-\br'|,\nu)$, so that $\langle \br |R(\mu)|\br'\rangle=
R(|\br-\br'|,\mu)$ depend only on the distance $|\br-\br'|$. Hence we can set
$\br'={\bf 0}$ without loss of generality.
The Fourier transform
\be
\hat{R}(p,\mu)=\int d\br e^{i\bp\cdot\br}R(\br,\mu),\quad p=|\bp|
\label{4.3}
\ee
gives the  distribution of momentum $\hbar \bp$ of an interacting Bose
particle. It is determined by equation  (\ref{4.2})
in the Fourier representation
\be
\hat{R}(p,\mu)=\hat{F}(p,\mu-a\rho(\mu))
\label{4.4}
\ee

\subsection{Grand-canonical density}

We first consider the diagonal part of the density matrix (i.e. the
particle density)
up to terms of
order $\gamma^{3}$
\be
\rho (\mu) = \rho_{mf}(\mu) + \gamma^{3}\rho^{(1)}(\mu ) \label{20}
\ee
where $\rho_{mf}$ is the solution of equation (\ref{8}). Using (\ref{5}) and
the $\gamma$-expansion (\ref{6}) of function $F^{\gamma}$  we
thus find
\be
\rho_{mf}(\mu) + \gamma^{3}\rho^{(1)}(\mu ) = \rho_{0}[\mu -a\rho_{mf}(\mu)
- a\gamma^{3}\rho^{(1)}(\mu)] + \gamma^{3}F^{(1)}[\mu -a\rho_{mf}(\mu)]
\label{21}
\ee
\[ = \rho_{0}[\mu -a\rho_{mf}(\mu)] - \gamma^{3}a\rho^{(1)}(\mu )
\rho'_{0}[\mu -a\rho_{mf}(\mu)] + \gamma^{3}F^{(1)}[\mu -a\rho_{mf}(\mu)] \]
Owing to equation (\ref{8}) the above relation yields the formula
\be
\rho^{(1)}(\mu) = \frac{F^{(1)}[\mu -a\rho_{mf}(\mu)]}
{1 + a\rho'_{0}[\mu -a\rho_{mf}(\mu)] }   \label{22}
\ee
where $F^{(1)}$ is given by (\ref{19}).

  The density correction $\rho^{(1)}(\mu)$ involves functions of the argument
\be
\nu(\mu) = \mu - a\rho_{mf}(\mu)    \label{23}
\ee
Notice that on one hand
\[  \frac{\partial\nu}{\partial\mu} = 1 - a \rho'_{mf}(\mu) ,\]
and on the other hand the mean field equation (\ref{8}) implies
\[ \frac{\partial\nu}{\partial\mu} = \frac{\rho'_{mf}(\mu)}{\rho'_{0}(\nu)} \]
so that
\be
\frac{\partial\nu}{\partial\mu} = \frac{1}{1 + a \rho'_{0}(\nu)}  \label{24}
\ee
The density correction (\ref{22}) can be thus rewritten in the form
\be
\rho^{(1)}(\mu) = F^{(1)}[\nu(\mu)]\frac{\partial\nu}{\partial\mu} \label{25}
\ee
which in view of the structure of equation (\ref{19}) finally yields
\be
\rho^{(1)}(\mu) = \frac{\partial g(\nu(\mu))}{\partial \mu}  \label{26}
\ee
where function $g(\nu)$ has been defined in (\ref{199}).

This correction to the grand-canonical density can also be studied at low
density and in the critical region.
If $\nu\sim\mu\to-\infty$ (i.e. low density) one finds that
$\rho^{(1)}(\mu) \sim A(\beta) e^{2\beta\mu}$
where, as for the pressure, the sign of the coefficient $A(\beta)$ depends
if effects of Bose statistics dominate
classical corrections or not. Extrapolating the formula (\ref{26}) to the
neighborhood
of the critical mean field chemical potential $\mu_{mf,c}=a\rho_{0,c}$, one
deduces first
from the mean field equation that $\nu(\mu)\sim
-C(\beta)(\mu_{mf,c}-\mu)^{2}, C(\beta)>0,$ as
$\mu\to \mu_{mf,c} \,(\mu<\mu_{mf,c})$.
Then analyzing the $\bk$-integrals as in subsection 3.3 one sees that
$\rho^{(1)}(\mu)$ diverges to  $-\infty$
as $\mu\to \mu_{mf,c}$ (for $\hat{V}(k)\sim bk^{-\eta}$,
$\rho^{(1)}(\mu)\sim -(\mu_{mf,c}-\mu)^{-(1+3/2\eta)}$
as for the pressure). Thus, taking $\mu$ close to $\mu_{mf,c}$ and $\gamma$
small enough, the density decreases when
fluctuations are taken into account.
This is illustrated in Fig.2.
\begin{figure}
\includegraphics{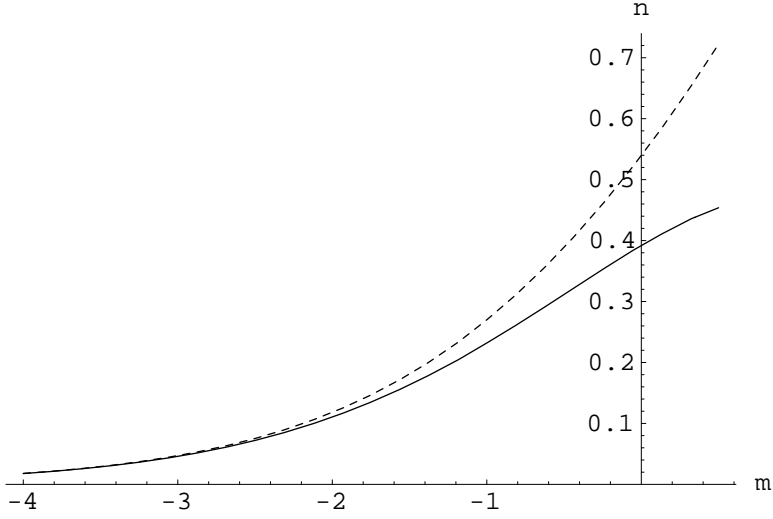}
\caption{Dimensionless density (\ref{21}): $ n = \rho (2\pi
\lambda^2)^{3/2}$ as function of
 the chemical potential $m=\mu\beta$ for the binary interaction (\ref{F1}).
The dashed line
represents the mean-field prediction which overestimates the density. We
put here
$\alpha \equiv a\beta/(2\pi\lambda^2)^{3/2}=1.5$, $\alpha
(\gamma\lambda/r_{0})^{3}=4\sqrt{2/\pi }$.
The density correction diverges at $m=m_{0,c}=a\rho_{0,c}=3.93$. }
\end{figure}

\subsection{The momentum distribution}

The momentum distribution $\hat{R}(p,\rho)$ at order $\gamma^{3}$,
considered as a function of the density
\be
\hat{R}(p,\rho)=\hat{R}^{(0)}(p,\rho)+\gamma^{3}\hat{R}^{(1)}(p,\rho)
\label{4.5}
\ee
is found from the relation (\ref{4.4})
by using  the $\gamma$-expansion of $\hat{F}^{\gamma}$ which follows from
(\ref{A.19}) an from the relation
$\nu(\mu_{mf}(\rho))=[\mu_{mf}-a\rho(\mu)]_{\mu=\mu_{mf}}(\rho)=\mu_{0}(\rho)$.
We observe that at the dominant order (see (\ref{A.10}),(\ref{A.11}))
\be
\hat{R}^{(0)}(p,\rho)= n_{0}(p,\nu)|_{\nu=\mu_{0}(\rho)}
\label{4.6}
\ee
is the free Bose momentum distribution at density $\rho$.
Proceeding as in (\ref{32})-(\ref{34}) one finds
\be
\hat{R}^{(1)}(p,\rho)=\hat{F}^{(1)}(p,\mu_{0}(\rho))-\left[\frac{n'_{0}(p,\nu)F^{(1)}(\nu)}
{\rho'_{0}(\nu )}\right]_{_{\nu=\mu_{0}(\rho)}}
\label{4.7}
\ee
Inserting the expression (\ref{A.19}) for $\hat{F}^{(1)}(p,\nu)$ and for
$F^{(1)}(\nu)=\int d\bp\hat{F}^{(1)}(p,\nu)/(2\pi)^{3}$
one obtains the final result
\be
\hat{R}^{(1)}(p,\rho)=c(\rho,\beta)
\left[\frac{\rho''_{0}(\nu)}{\rho'_{0}(\nu)}n'_{0}(p,\nu)-n''_{0}
(p,\nu)\right]_{\nu=\mu_{0}(\rho)}
\label{4.8}
\ee
where \[ c(\rho,\beta)= \frac{1}{2\beta}
\int\frac{d\bk}{(2\pi)^{3}}\(\frac{\hat{V}(k)}{1+\hat{V}(k)\rho'_{0}(\mu_{0}
(\rho))} \)  \]
Notice that
\be
\int\frac{d\bp}{(2\pi)^{3}}\hat{R}^{(1)}(p,\rho)=0
\label{4.8a}
\ee
as requested by the fact that the integrals
\be
\int\frac{d\bp}{(2\pi)^{3}}\hat{R}(p,\rho)=\int\frac{d\bp}{(2\pi)^{3}}
\hat{R}^{(0)}(p,\rho)=\rho
\label{4.8b}
\ee
are fixed by the total density.

Some properties of $\hat{R}(p,\rho)$ can be derived from those of the free Bose
distribution
\[ n_{0}(p,z)=\frac{z}{\exp(\lambda^{2} p^{2}/2)-z}  \]
where $z=e^{\beta\nu}$, $0\leq z<1$,  is the activity parameter.
One sets
\beq
n_{1}(p,z)&=&\beta^{-1}n'_{0}(p,\nu)=n_{0}(p,z)(1+n_{0}(p,z))\nonumber\\
n_{2}(p,z)&=&\beta^{-2}n''_{0}(p,\nu)=n_{1}(p,z)(1+2n_{0}(p,z))
\label{4.9}
\eeq
Then
\be
\hat{R}(p,z)=n_{0}(p,z)+\epsilon[n_{1}(p,z)r(z)-n_{2}(p,z)]
\label{4.10}
\ee
with
\be
r(z)=\frac{\rho''_{0}(\nu)}{\beta\rho'_{0}(\nu)}= \frac{\int d\bp\, n_{2}(p,z)}
{\int d\bp\, n_{1}(p,z)}
\label{4.11}
\ee
The parameter $\epsilon  = \gamma^{3}c(\rho,\beta)\beta^{2}$ in
(\ref{4.10}) incorporates
all the $p$-independant factors (see eq.(\ref{4.8})).
Since $\epsilon$ is proportional to $\gamma^{3}$ it can be chosen as small
as one wishes at any density $\rho<\rho_{0,c}$.
We observe that
\be
n_{2}(p,z)\leq n_{1}(p,z)\(1+\frac{2z}{1-z}\) =
n_{1}(p,z)\(\frac{1+z}{1-z}\)
\label{4.12}
\ee
implying
\beq
r(z)&\leq& \frac{1+z}{1-z}\nonumber\\
\left[n_{1}(p,z)r(z)-n_{2}(p,z)\right]_{p=0}&=&\frac{z}{(1-z)^{2}}\left[r(z)
- \frac{1+z}{1-z}\right]\leq 0
\label{4.13}
\eeq
One concludes that the distribution at $p=0$ in presence of the interaction
is always less than the free value $z/(1-z)$.
Evaluating the correction as $p\to\infty$ gives
\be
[n_{1}(p,z)r(z)-n_{2}(p,z)]\sim z\exp(-\lambda^{2} p^{2}/2)(r(z)-1)\geq 0
\label{4.14}
\ee
since $n_{0}(p,z)\sim n_{1}(p,z)\sim n_{2}(p,z)\sim z\exp(-\lambda^{2}
p^{2}/2),\;p\to\infty$,
and for all $p$ and $z$, $ n_{2}(p,z)\geq n_{1}(p,z)$, implying $r(z)\geq 1$.
Thus the momentum distribution is flattened and broadened by repulsive
interactions, as illustrated in Fig.3.
\begin{figure}
\includegraphics{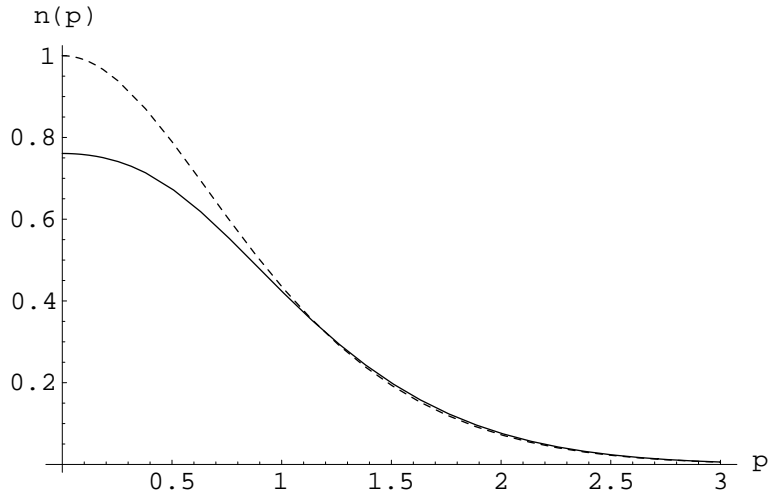}
\caption{Momentum density (\ref{4.10}): $ n(p) = \hat{R}(p,0.5)$ as function of
 momentum p. We put $\lambda=1$ and $\epsilon = 0.9$ . The dashed line
represents the ideal gas Bose distribution.}
\end{figure}

\section{Concluding remarks}

One can make the following points:

(i) Divergences occuring at the critical values of the free gas reflect the fact
that the order $\gamma^{3}$ of the
correction is not adequate there. Since we know from rigorous work that the
exact pressure $P_{\gamma}$ converges
to the mean field pressure $P_{mf}$ for all values of the thermodynamical
parameters \cite{SZ}
one may conclude that the rate of approach to mean field at critical values
of the free gas is of the order $\gamma^{\delta}$
for some $0< \delta <3$. The determination of this exponent is an open
problem which necessitates a more thorough
analysis of Mayer series.

(ii) In our analysis, the critical values of density and of temperature
are still those of the free gas. One
could expect that for non-zero but small $\gamma$
the critical mean field chemical potential $\mu_{mf,c}=a\rho_{0,c}$ and
density $\rho_{0,c}$ will be sligthly displaced.
Such an information cannot been obtained here, but requires, for instance,
a deeper non perturbative understanding of the
behaviour of the partition of a single polymer (\ref{10}) as also discussed
in Section V.A  of \cite{MP}.

(iii) The results of the paper enable nevertheless to explore the immediate
vicinity of the critical mean field
values by taking $\gamma$ sufficiently small (provided that our formulae
keep their validity up to this point).
Then qualitative statement could be established, as the fact that the exact
pressure should be higher  than that predicted in this region by the mean
field .
Another conclusion from our analysis is that approaching the critical
mean field chemical potential $\mu_{mf,c}=a\rho_{0,c}$ along an isotherm
one finds
the interacting gas at a density lower than that appearing in the mean
field approach.\\[0.5cm]
{\bf Acknowledgments}

\noindent JP greatly acknowledges the hospitality at the \'Ecole
Polytechnique F\'ed\'erale de Lausanne and Ph.M. hospitality at the University of Warsaw,
places where this research has been performed.
JP acknowledges the financial support for [1] and for the 
present work by the the Committee for Scientific Reseach (KBN, Poland), 
grant 2P03B00823.   

\appendix
\section{Appendix}
\subsection{Polymer representation of the one-body reduced density matrix}

The representation of the off-diagonal reduced density matrix
$\langle \br |R(\mu)|\br'\rangle$ in the formalism of random polymers
described in Section 2
of \cite{MP} requires the consideration of  open Brownian paths (open polymers)
in the time interval
$0\leq s \leq q $ with extremities at
$\br$ and $\br'$. In the Brownian bridge notation they are parametrized as
\be
\(1-\frac{s}{q}\)\br+\frac{s}{q}\br'+\lambda\bX(s),\quad 0\leq s \leq q .
\label{A.1}
\ee
where $\bX(s)$ is a closed path distributed with the
Brownian bridge measure (eq. (18) in \cite{MP} ; here and in the sequel we use
the same notations as in
\cite{MP}). The open polymer $\cl_{\br,\br'}=(\br,\br', q, \bX(s)),\; 0\leq s
\leq q$ is
characterized by its end points $\br,\br'$, the number $q$ of particles
belonging to it and its random shape
$\bX(s)$. By a slight generalisation of the analysis that led to the "magic
formula" (eq. (14) in \cite{MP})
one obtains the density for an open polymer immersed in a grand canonical
ensemble of closed polymers
as
\beq
&&\rho_{{\rm op}} (\cl_{\br,\br'})=\nonumber\\
&&\frac{1}{\Xi_{\Lambda}}\sum_{n=1}^{\infty}\frac{1}{(n-1)\,!}
z(\cl_{\br,\br'})\int\prod_{i=2}^{n} d\cl_{i}z(\cl_{i})\exp(-\beta
U(\cl_{\br,\br'},\cl_{2},\ldots,\cl_{n})).\nonumber\\
\label{A.2}
\eeq
where $\Xi_{\Lambda}$ is the partition function.
If $\br=\br'$ this reduces to the the loop density $\rho_{{\rm loop}}$
studied in \cite{MP}
\footnote{The quantity $-k_{B}T\log \rho_{{\rm op}} (\cl_{\br,\br'})$ can be
interpreted as the excess grand potential
when a fixed open polymer $\cl_{\br,\br'}$ is introduced in the system of
closed polymers.}.
In view of its form (\ref{A.2}) $\rho_{{\rm op}} (\cl_{\br,\br'})$ has the
Mayer expansion presented in
eqs.(29)-(31) of \cite{MP} in terms of Ursell functions. The only difference is
that the argument
of the root point $1=\cl_{\br,\br'}$ has to be identified with the open
polymer (\ref{A.1}).
Polymer interactions, effective activities and Mayer bonds are the same as
in \cite{MP}
except at the root point,
where the interaction between a loop $\cl_{j}=(\bR_{j},q_{j}, \bX_{j})$
with an open polymer is
\begin{eqnarray}
&& V(\cl_{\br,\br'},\cl_j)=\nonumber\\
&& \int_0^{q}ds\int_0^{q_{j}}ds_{j}{\hat \delta}
(s-s_{j})V\(\(1-\frac{s}{q}\)\br +\frac{s}{q}\br' + \lambda
\bX(s)-\bR_j-\lambda\bX_{j}(s_j)\) \nonumber\\
\label{A.3}
\end{eqnarray}
and the self-energy of the open polymer is
\beq
 U(\cl_{\br,\br'})&=&
\frac{1}{2}\int_0^{q}
ds_1\int_0^{q}ds_2{\hat \delta}(s_1-
s_2)V\(\frac{s_{1}-s_{2}}{q}(\br-\br')+\lambda(\bX(s_1)-\bX(s_2))\)\nonumber
\\
 &-&\frac{1}{2}qV(0)
\label{A.4}
\eeq
Finally, to obtain the one-body density matrix $\langle \br
|R(\mu)|\br'\rangle$, one has to integrate on the internal degrees
of freedom of the open polymer
\be
\langle \br|R(\mu)|\br'\rangle=\sum_{q=1}^{\infty}q\int D_{q}(\bX)
\exp\(-\frac{|\br-\br'|^{2}}{2q\lambda^{2}}\)\rho_{{\rm op}} (\cl_{\br,\br'})
\label{A.5}
\ee
The $q$ factor takes into account the presence of $q$ particles
in the open polymer and the additionnal Gaussian
comes from the Brownian Wiener weight for a path starting in $\br$ at time
$s=0$ and ending at $\br'$ at time $s=q$.
From now on the analysis of the Mayer series  can be performed exactly
along the same lines as that given in \cite{MP}
with the following results.

\vspace{2mm}

\noindent (i) The Mayer series representing $\langle \br
|R(\mu)|\br'\rangle$ converges for $\mu$ sufficiently negative
$(\mu<-a\rho_{0,c})$; for a scaled potential $V_{\gamma}(\br)$, the
convergence is uniform with respect to $\gamma$.

\vspace{2mm}

\noindent (ii) Let $I(\cl_{\br,\br'})$ be the value of the sum of all
multiply connected graphs with root point
$\cl_{\br,\br'}$, and define the function of the chemical potential $\nu$
\be
F_{\br,\br'}(\nu) = \sum_{q=1}^{\infty}q\int
D_{q}(\bX)\exp\(-\frac{|\br-\br'|^{2}}{2q\lambda^{2}}\)I(\cl_{\br,\br'}).
\label{A.6}
\ee
Then the reduced density matrix is given by
\be
\langle \br |R(\mu)|\br'\rangle = F_{\br,\br'}(\mu-a\rho(\mu)).
\label{A.7}
\ee
where the density $\rho(\mu)$ solves the self-consistent equation (\ref{5}).

\subsection{The $\gamma^{3}$ correction}

For a scaled potential ($\gamma \ll 1$), the function $F^{\gamma}(\br,
\nu)$ takes the asymptotic form
\be
F^{\gamma}(\br,\nu)=F_{tree}^{0}(\br,\nu)+\gamma^{3}[F_{tree}^{1}(\br,\nu)+F
_{ring}^{1}(\br,\nu)] + o(\gamma^{3})
\label{A.8}
\ee
as in (\ref{6}) where $F_{tree}^{(0)}(\br,\nu)$ represents the sum of zero
order contributions coming from the tree diagrams.
The sum of tree diagrams yields
\be
F_{tree}^{\gamma}(\br,\nu)=\frac{1}{(2\pi\lambda^{2})^{3/2}}\sum_{q=1}^{\infty }
\frac{\exp\(\beta\nu
q-\frac{r^{2}}{2\lambda^{2}q}\)}{q^{3/2}}\kappa^{\gamma}(q),quad r=|\br|
\label{A.9}
\ee
so that
\beq
F_{tree}^{(0)}(\br,\nu)&=&\lim_{\gamma\to 0}F^{\gamma}_{tree}(\br,\nu)
=\frac{1}{(2\pi\lambda^{2})^{3/2}}\sum_{q=1}^{\infty }
\frac{\exp\(\beta\nu q-\frac{r^{2}}{2\lambda^{2}q}\)}{q^{3/2}}\nonumber\\
&=&\int \frac{d\bk}{(2\pi)^{3}} e^{-i\bk\cdot\br}n_{0}(k,\nu)
\equiv\rho_{0}(\br,\nu)
\label{A.10}
\eeq
is nothing else than the off-diagonal reduced density matrix
$\rho_{0}(\br,\nu)= \langle \br |R_{0}(\nu)|{\bf 0}\rangle$
of the free gas at chemical potential $\nu~$. Indeed,
$F_{tree}^{(0)}(\br,\nu)$ is the Fourier
transform of the Bose occupation number density
$n_{0}(k,\nu)$
\be
n_{0}(k,\nu)=\frac{1}{\exp\(\frac{(\lambda |k|)^{2}}{2}-\beta\nu\)-1}.
\label{A.11}
\ee
The $\gamma^{3}$ correction is obtained in expanding the partition function
$\kappa^{\gamma}(q)$ of a single polymer in
$F_{tree}^{\gamma}(\br,\nu)$  as done in eqs (\ref{10}) to (\ref{14})
\beq
F_{tree}^{(1)}(\br,\nu)&=&\frac{\beta
V(0)}{2}\(f^{(1)}(\br,\nu)-f^{(2)}(\br,\nu)\),\quad {\rm with}\nonumber\\
f^{(k)}(\br,\nu)&=&\frac{1}{\beta^{k}}\frac{\partial^{k}}{\partial\nu^{k}}\rho_{0}(\br,\nu).
\label{A.12}
\eeq
Clearly $f^{(k)}(\br,\nu)|_{\br=0}=f^{(k)}(\nu)$ defined in (\ref{15}).
We now perform the ring summation to determine $F_{ring}^{(1)}(\br,\nu)$ .
The contribution to $I_{ring}(\cl_{\br,\br'})$
of a ring with one root point $\cl_{\br,\br'}$, $n$ integrated vertices
$\cl_{j}$
and $n+1$ linarized bonds $(-\beta V)$ is
\beq
&&\frac{1}{2}z(\cl_{\br,\br'})\int d\cl_{1}\cdots\int d\cl_{n}(-\beta
V(\cl_{\br,\br'},\cl_{1}))\nonumber\\
&\times&\left[\prod_{j=1}^{n-1}z(\cl_{j})(-\beta V(\cl_{j},\cl_{j+1}))
\right](-\beta V(\cl_{n},\cl_{\br,\br'}))
\label{A.13}
\eeq
The factor $1/2$ is the symmetry factor of the graph. Introducing the
scaled potential $\gamma^{3}V(\gamma\br)$
and changing the $n$ spatial integration variables $\bR_{j}$ to
$\gamma\bR_{j}$ will produce an overall factor
$\gamma^{3}$ in (\ref{A.13}) since there are $n+1$ bonds. At order
$\gamma^{3}$, we are thus entitled to neglect
the quantum fluctuation part $\gamma \lambda \bX(s_{j})$ in the arguments
of the potential
(use dominated convergence and the fact that the Gaussian measures
$D_{q_{j}}(\bX_{j})$ are normalized).
For the same reason we neglect the $\gamma$ dependence in the activities.
Within this approximation
we take the bonds and the vertices equal to
\beq
V_{\gamma}(\cl_{\br,\br'},\cl_j)&= & \gamma^{3}q_{j}\int_0^{q}ds
V\(\(1-\frac{s}{q}\)\gamma\br
+\frac{s}{q}\gamma\br'-\gamma\bR_{j}\)\nonumber\\
V_{\gamma}(\cl_{j},\cl_{j+1})&= &  q_{j}q_{j+1}
\gamma^{3}V(\gamma\bR_{j}-\gamma\bR_{j+1})\nonumber\\
z(\cl_{\br,\br'})&= &z(\cl_{j})=\frac{e^{\beta\nu
q}}{q(2\pi\lambda^{2}q)^{3/2}}\equiv z^{(0)}(q)
\label{A.14}
\eeq
At this point we keep the $\gamma$
parameter in combinations $\gamma\br$ and $\gamma\br'$ since $\br,\;\br'$
can be  large.
Hence after the change of variables
$\bR_{j}\rightarrow \gamma\bR_{j}$ (\ref{A.13}) becomes
\beq
&&\frac{\gamma^{3}}{2}z^{(0)}(q)\sum_{q_{1},\ldots,q_{n}}\int
d\bR_{1}\cdots d\bR_{n}\left[-\beta
\int_0^{q}dsV\(\(1-\frac{s}{q}\)\gamma\br
+\frac{s}{q}\gamma\br'-\bR_{1}\)\right]\nonumber\\
&&\(\prod_{j=1}^{n}q_{j}^{2}z^{(0)}(q_{j})\)
\(\prod_{j=1}^{n}(-\beta
V(\bR_{j}-\bR_{j+1}))\)\int_0^{q}ds'V\(\bR_{n}-\(1-\frac{s'}{q}\)\gamma\br
-\frac{s'}{q}\gamma\br'\)\nonumber\\
\label{A.15}
\eeq
Introducing the Fourier transform $\hat{V}(k)$ and using the convolution
theorem
yields
\beq
\frac{\gamma^{3}}{2}z^{(0)}(q)(f^{(1)}(\nu))^{n}\int_{0}^{q}ds\int_{0}^{q}ds
'\int
\frac{d\bk}{(2\pi)^{3}}\exp\left[i\gamma
\bk\cdot(\br-\br')\(\frac{s-s'}{q}\)\right] (-\beta
\hat{V}(k))^{n+1}\nonumber\\
\label{A.16}
\eeq
where $f^{(1)}(\nu)=\sum_{q=1}^{\infty}q^{2}z^{(0)}(q)$ is the function
(\ref{15}). This has to be summed on $n=1,2,\ldots$
and integrated over the internal degrees of freedom of the root point
according to (\ref{A.6}).  One finds
\beq
&&F_{ring}^{(1)}(\br-\br',\nu)=\frac{1}{2}\sum_{q=1}^{\infty}qz^{(0)}(q)
\exp\(-\frac{|\br-\br'|^{2}}{2q\lambda^{2}}\)
\nonumber\\
&&\int\frac{d\bk}{(2\pi)^{3}}\int_{0}^{q}ds\int_{0}^{q}ds'\exp\left[i\gamma\
\bk\cdot(\br-\br')\(\frac{s-s'}{q}\)\right]
\frac{(\beta\hat{V}(k))^{2}f^{(1)}(\nu)}{1+\beta\hat{V}(k)f^{(1)}(\nu)}\nonumber\\
\label{A.17}
\eeq
Finally, taking (\ref{A.14}) into account,  the Fourier transform
$\hat{F}^{(1)}(p, \nu)=\int d\br e^{i\bp\cdot\br}F_{ring}^{(1)}(\br,\nu)$
at wave number $\bp$ is found to be
\beq
&&\hat{F}_{ring}^{(1)}(p, \nu)=\frac{1}{2}\sum_{q=1}^{\infty}q e^{\beta\nu
q}\int_{0}^{q}ds\int_{0}^{q}ds'\int\frac{d\bk}{(2\pi)^{3}}\nonumber\\
&& \exp\left[-\frac{\lambda^{2}}{2}\left|p + \gamma
k\(\frac{s-s'}{q}\)\right|^{2}q
\right]
\frac{(\beta\hat{V}(k))^{2}f^{(1)}(\nu)}{1+\beta\hat{V}(k)f^{(1)}(\nu)}
\nonumber\\
&=&\frac{1}{2}\hat{f}^{(2)}(p,\nu)\int\frac{d\bk}{(2\pi)^{3}}
\frac{(\beta\hat{V}(k))^{2}f^{(1)}(\nu)}
{1+\beta\hat{V}(k)f^{(1)}(\nu)}+o(\gamma )
\label{A.18}
\eeq
As a consequence of the Gaussian weight occuring for off-diagonal elements
$\br \neq \br'$
the remaining $\gamma$ dependance has been neglected in (\ref{A.18}).
$\hat{f}^{(2)}(p,\nu)$ is the Fourier transform of the function (\ref{A.12}).
We note that
\[ \int
\frac{d\bp}{(2\pi)^{3}}\hat{F}_{ring}^{(1)}(p,\nu)=F_{ring}^{(1)}(\nu) \]
reduces to the formula (\ref{16}) used in Section 2.

The final result for the $\gamma^{3}$ correction is obtained by adding the
tree contribution (\ref{A.12}) to (\ref{A.18})
\beq
\hat{F}^{(1)}(p,\nu)&=&\hat{F}_{tree}^{(1)}(p,\nu)+\hat{F}_{ring}^{(1)}(p,\nu)
\nonumber\\
&=&\frac{V(0)}{2}n_{0}'(p,\nu)
-\frac{1}{2\beta}n_{0}''(p,\nu)\int\frac{d\bk}{(2\pi)^{3}}
\; \frac{\hat{V}(k)}{1+\hat{V}(k)\rho_{0}'(\nu)}
\label{A.19}
\eeq
where we have expressed all quantities in terms of the density and the Bose
momentum  distribution (\ref{A.11}) of the free gas.

\end{document}